\documentclass[prl,twocolumn,aps,floatfix,nofootinbib,natbib,superscriptaddress]{revtex4-2}

\usepackage{pslatex,graphicx,dcolumn,bm,natbib,amssymb,amsmath,color,mathtools,mathrsfs,eufrak}

\usepackage[normalem]{ulem}
\usepackage[utf8]{inputenc}
\usepackage[T1]{fontenc}
\usepackage{hyperref}
\usepackage[english]{babel}
\usepackage{blindtext}
\usepackage{booktabs}
\newcommand{\beq}{\begin{equation}}
\newcommand{\eeq}{\end{equation}}

\newcommand{\beqa}{\begin{eqnarray}}
\newcommand{\eeqa}{\end{eqnarray}}

\newcommand{\ka}{$\kappa_A$}
\newcommand{\ts}{$\gamma_{\text{true}} \ $}

\newcommand{\be}{\begin{equation}}
\newcommand{\ee}{\end{equation}}
\newcommand{\bea}{\begin{eqnarray}}
\newcommand{\eea}{\end{eqnarray}}

\begin{document}

\title{Shear-driven solidification and nonlinear elasticity in  epithelial tissues}

\author{Junxiang Huang}
\affiliation{Department of Physics, Northeastern University, MA, USA}
\author{James O. Cochran}
\affiliation{Department of Physics, Durham University, Science Laboratories, South Road, Durham DH1 3LE, UK}
\author{Suzanne M. Fielding}
\affiliation{Department of Physics, Durham University, Science Laboratories, South Road, Durham DH1 3LE, UK}
\author{M. Cristina Marchetti}
\affiliation{Department of Physics, University of California, Santa Barbara, CA, USA}
\author{Dapeng Bi}
\affiliation{Department of Physics, Northeastern University, MA, USA}

\begin{abstract}
 Biological processes, from  morphogenesis to tumor invasion, spontaneously generate shear stresses inside living tissue. The mechanisms that govern the transmission of mechanical forces in  epithelia and the collective response of the tissue to bulk shear deformations remain, however, poorly understood.  Using  a minimal cell-based computational model, we investigate the constitutive relation of confluent tissues under simple shear deformation. We show that an initially undeformed fluid-like tissue    acquires finite rigidity above a critical applied strain.   This is akin to the shear-driven rigidity  observed in other soft matter systems. Interestingly,  shear-driven rigidity can be understood by a critical scaling analysis in the vicinity of the second order  critical point that governs the liquid-solid transition of the undeformed system. We further show that a solid-like tissue responds linearly only to small strains and but then switches to  a nonlinear response at larger stains, with substantial stiffening. Finally, we propose a mean-field formulation for cells under shear that offers a simple physical explanation of shear-driven rigidity and nonlinear response in a tissue. 
\end{abstract}
\maketitle

 Monolayers of tightly connected cells provide essential physical barriers and filters to all organs \emph{in vivo}.
The tight connections between cells
allow the tissue
to resist external deformation and withstand stress, while maintaining its integrity. At the single cell level, researchers have used a broad  repertoire of experimental techniques\cite{Fabry_review_nonlinear_cell_mech,guck_optical_stretcher,AFM_review_2015,AFM_review_2020,Fujii_AFM,Serwane_droplet_NM} to reveal a rich  mechanical behavior,
including power-law rheology\cite{Hoffman10259} and stress stiffening\cite{fernandez2006master}. At the mesoscopic level, traction force microscopy has allowed the mapping of intercellular forces\cite{Schwarz_TFM_review,Tambe_TFM,Butler_TFM_review}, revealing a rough stress landscape, with spatial fluctuations correlated over several cells\cite{Mertz_PNAS,Mertz_PRL,Trepat_nphys_forces,Kim_propulsion_NM_2013}.

There is increasing consensus that mechanical deformations can directly influence {collective} cell behavior\cite{Getsios_rev,Ingber_rev,Martino_rev,Zhang_signaling_review,Das_NCB_2015} and play a central role in driving developmental processes\cite{Hayes_dorsal_closure_review,Machado2015,ANDREW201034,Etournay_elife,Guirao_elife,Lecuit_force_review,tetley2019tissue,Wang13541},
physiology\cite{Fisher_endothelial_shear,trepat_fredberg_stretch_nature_2007,riveline_elife_colony_elongate, Trepat_nphys_forces,tambe2011collective,Das_NC_2018}, 
and tumor progression\cite{valerie_cancer_review,wirtz2011physics,Jain_cancer_review}. 
Experiments\cite{trepat_fredberg_stretch_nature_2007,Harris_PNAS_stretch,khalilgharibi2019stress,Pruitt_shear_elife} have shown that epithelial monolayers {respond nonlinearly to external mechanical stretch, with observed}  
stress-stiffening and even fracturing. Similar behavior has been observed in tissues  deformed by internal active motile forces\cite{prakash2021motility} and in curved epithelial sheets enclosing an expanding lumen\cite{latorre2018active}. Importantly, these experimental studies have typically focused on probing the behavior of solid-like tissue, where cells do not spontaneously exchange neighbors. 
 On the other hand, the last decade has seen a surge of evidence {demonstrating that living tissue can spontaneously undergo transitions} 
between a solid-like (jammed) state and a fluid-like (unjammed) state.
\cite{
Park_NMAT_2015,
Garcia2015,
Oswald2017,
paul2017cancer,
malinverno2017endocytic,
atia2018geometric,
mongera2018fluid,
Ilina2020,
Huebner2020,
mitchel_ncomm_2020,
Petridou_percolation,
lin2021energetics,
haiqian_pnas_2021,
De_Marzio_sci_adv}.
Despite {its} fundamental importance and direct relevance to biology, the response of a cell collective  to mechanical deformation \emph{at the tissue level} remains poorly understood, especially in the vicinity of the tissue solid-fluid transition. 

A growing number of theoretical studies has begun to address this gap. Various groups have used vertex-based models\cite{Nagai_PMB_2001,Farhadifar_CB_2007} to simulate 
the linear\cite{Rastko_2021} and nonlinear\cite{Popovic_2021,Duclut_nonlinear,PicaCiamarra_rheology} rheology of a tissue under steady shear.  The effects of active tension fluctuations\cite{krajnc2021,Duclut_nonlinear} and cell division\cite{Li_Bo_shear_work} have been explored. An earlier study\cite{Merzouki_vm_strain} has showed that the vertex model exhibits a nonlinear mechanical response qualitatively similar to experiments\cite{Harris_PNAS_stretch}.
Despite this growing body of work, to date there is no systematic study of the mechanical response of an amorphous epithelial tissue near the solid-fluid transition. 

Here we use a cell-vertex model to investigate the
 tissue response to externally imposed shear deformations.
 We show that a tissue  which is fluid-like when undeformed 
acquires rigidity {above}   a threshold value of the applied strain.  This is akin to the shear-driven rigidity of fiber networks and shear jamming in
 granular matter\cite{bi2011jamming}.  The {onset of}   shear-driven rigidity {in the liquid state} is  {characterized by a discontinuous jump in the tissue shear modulus}, and { the size of the jump } depends on {the} distance to the  second order liquid-solid critical point of the undeformed system. 
We find that nonlinear elasticity  becomes increasingly dominant  closer to the critical point, where the  mechanical response is completely nonlinear. This intrinsic critical nonlinearity was also demonstrated in recent work on a vertex models of regular polygons, where it was shown to arise from purely geometric constraints\cite{hernandez2021geometric}.
While Ref.\cite{hernandez2021geometric} focused on 
the response to infinitesimal perturbations, 
demonstrating the failure of linear elasticity, here we examine the nonlinear response in the presence of topological rearrangements that mediate plasticity.
We additionally extend the mean-field (MF) formulation of~\cite{hernandez2021geometric} 
to account for  the  emergence of {shear-induced} rigidity in the liquid state. The MF predicts exactly the nonlinear response and stress-stiffening exponents observed in the simulations.

\paragraph{Model.} We model a 2D cell layer using the Voronoi-based implementation\cite{Bi_PRX_2016,Li_Sun_Biophys} of the vertex model\cite{Farhadifar_CB_2007,PhysRevLett.123.058101,Li_PNAS_2018,yan_bi_prx,mitchel_ncomm_2020,das_t1_preprint}. 
Here,  the cell centers $\{\mathbf{r_i}\}$ are the degrees of freedom and their Voronoi tessellation  determine  the cellular structure\cite{Bi_PRX_2016}.  
\begin{figure}[htbp]
	\centering
	\includegraphics[width=1\columnwidth]{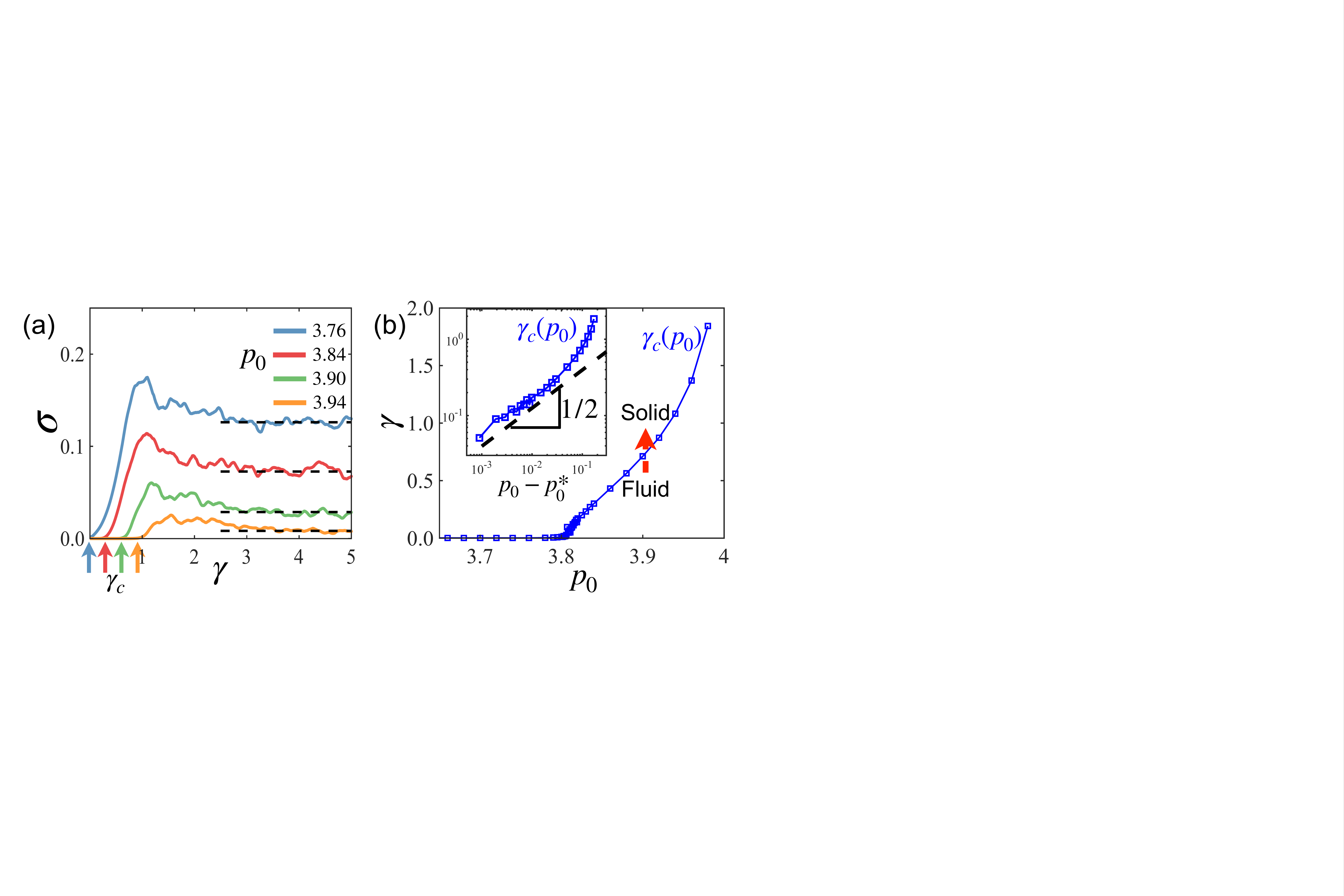}
	\caption{
	{\bf(a)} Stress-strain at different $p_0$ and $\kappa_A=0$. 
	An initially fluid-like tissue undergoes strain-driven rigidity above a critical threshold $\gamma_C$(location indicated by vertical arrows).  
	{\bf (b)} The critical strain $\gamma_C(p_0)$ defines a boundary that separates a fluid-like tissue from a solid-like tissue. Inset: $\gamma_c$ vs $p_0$ on log-log scale.
	}
	\label{fig_stress-strain}
\end{figure}
The mechanics of the cell layer is governed  by the energy function\cite{Staple_PJE_2010} 
$E= \sum_{i=1}^N \left[ K_A (A_i-A_0)^2+ K_P (P_i-P_0)^2 \right]$.
The first term,  quadratic in the cell areas $\{A_i\}$,  originates from the incompressibility of cell volume, giving rise to a 2D  area elasticity constant $K_A$ and preferred area $A_0$\cite{Farhadifar_CB_2007,Staple_PJE_2010}. 
The second term  quadratic in the cell  perimeters $\{P_i\}$  arises from the contractility of the cell cortex, with an elastic constant $K_P$\cite{Farhadifar_CB_2007}. Here $P_0$ is the target cell perimeter\cite{bi_nphys_2015}, representing the interfacial tension set by the competition between the cortical tension and the adhesion between  adjacent cells\cite{Staple_PJE_2010}.
In this work, we focus on the case where all cells have homogeneous single cell parameters $K_A, K_P, A_0, P_0$, while noting that the results {are easily} generalized to a tissue containing cell-to-cell heterogeneity\cite{PhysRevLett.123.058101} and are not qualitatively affected by this assumption. We choose $A_0=\bar{A}$, the mean cell area, which also serves as the length unit. The resulting non-dimensionalized energy is 
\be
E = \sum_{i=1}^N \kappa_A (a_i - 1)^2 + (p_i - p_0)^2,
\label{vm_energy}
\ee
with $\kappa_A=K_A\bar{A}/K_P$  the rescaled area elasticity. {Here} $p_0=P_0/\sqrt{\bar{A}}$ is a crucial model parameter called \emph{target cell shape index}. 
To study tissue response beyond the linear  regime\cite{yan_bi_prx}, we impose quasistatic simple shear  using Lees-Edwards boundary conditions\cite{Tildesley_book}.  Starting from a strain-free state ($\gamma=0$), the strain $\gamma$  is increased in increments of  $\Delta\gamma=2\times10^{-3}$, while   cell center positions {are subject} to an affine displacement 
$\Delta \mathbf{r_i} = \Delta\gamma \ y_i \ \hat{x}.$
Following each strain step,  Eq.\eqref{vm_energy} is relaxed using the FIRE algorithm\cite{Bitzek_PRL_Fire_2006} until all forces $\mathbf{F}_i \equiv -\partial {E} /\partial {\mathbf{r_i}}$ are vanishingly small ( $<10^{-14}$). For all results presented in this work, we used 84 random initial configurations and $N=400$ cells.

The unstrained tissue is known to exhibit a liquid-solid transition as a function of   $p_0$ \cite{bi_nphys_2015,yan_bi_prx,Sussman_Merkel}. When 
$p_0$  is below the \emph{critical cell shape index} $p_0^*=3.81$
 and $\kappa_A=0$ the unstrained tissue behaves as a rigid solid, with a finite \emph{linear-response} shear modulus
$
G_0 \equiv \lim_{\gamma\to0} {\partial \sigma}/{\partial \gamma}.
$
When $p_0\ge p_0^*$,  the unstrained tissue is fluid and $G_0=0$.
This solid-fluid transition at $\gamma=0$ 
is now well-understood in terms of a Maxwell constraint-counting approach\cite{yan_bi_prx,damavandi2021energetic} and as driven by geometric incompatibility\cite{bi_nphys_2015,yan_bi_prx,moshe2018geometric,Merkel_minimal_length_PNAS_2019,kupferman2020continuum}. 

\paragraph{ Nonlinear shear response.}
To  characterize the mechanical response at finite $\gamma$, we compute the tissue shear stress\cite{Ishihara_2012,Chiou_Shraiman_2012,Yang_PNAS_2017} 
$
\sigma = \sigma_{xy} \equiv L^{-2}\sum_{i<j} T_{ij}^x  l_{ij}^y,
$
where $\mathbf{l}_{ij}$ is the vector of the junction shared by cells $i,j$ {and $L$ is the simulation box size}. At each junction, the line tension vector is given by
 $\mathbf{T}_{ij} = \partial E / \partial \mathbf{l}_{ij}=2[(p_i-p_0) + (p_j-p_0)]  \hat{l}_{ij}$. 
\begin{figure}[htbp]
	\centering
	\includegraphics[width=1\columnwidth]{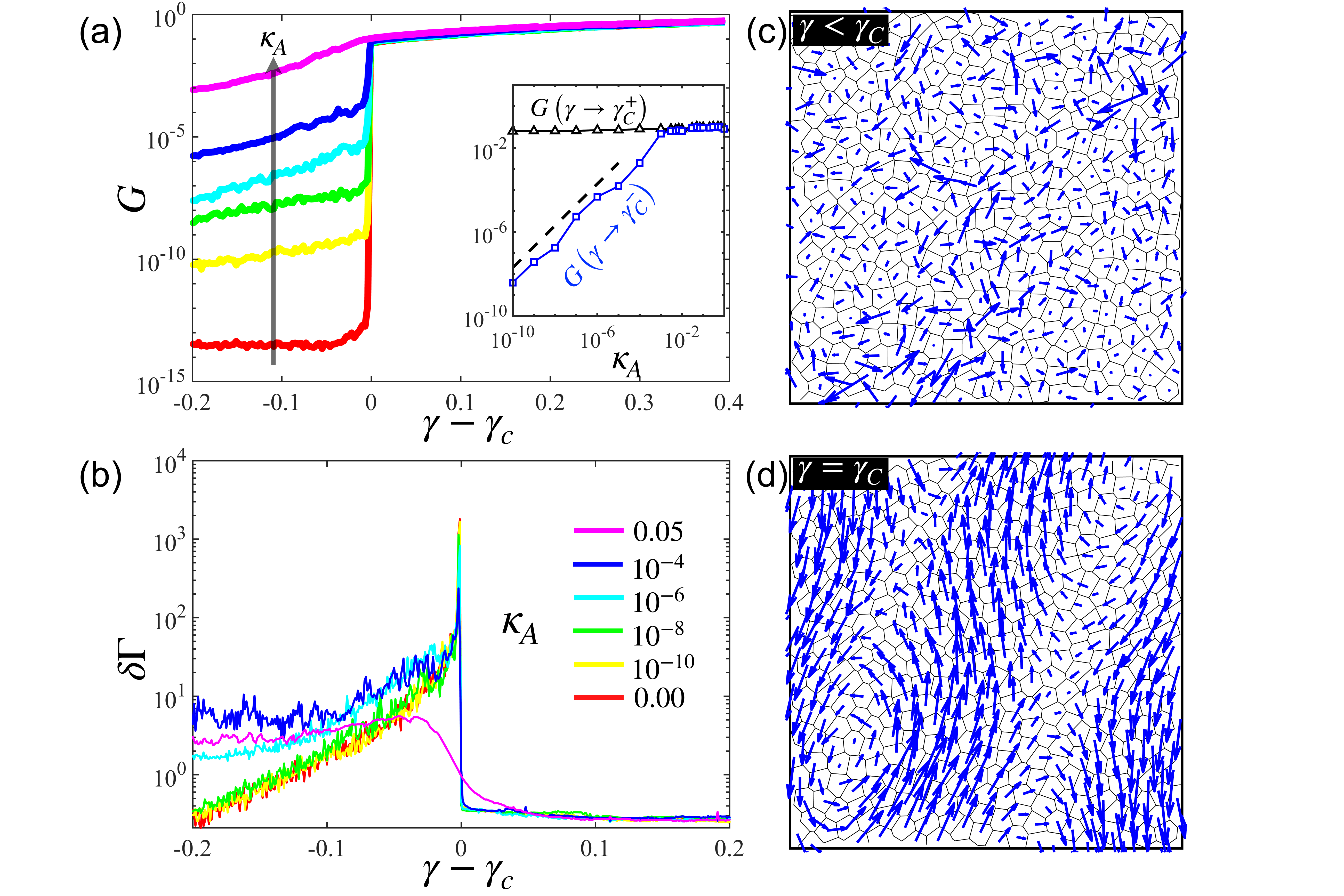}
	\caption{
	{\bf Strain-driven rigidity transition}
	{\bf(a)} The shear modulus $G$ near the onset of the strain-driven solidification for $p_0=3.84$ and   different area elasticities $\kappa_A=0,10^{-10},10^{-8},10^{-6}, 10^{-4}, 0.05$. Color legends provided in(b).
	{\bf Inset:}  $G$ immediately below and above the transition shows a gap that narrows with increasing \ka. The dashed line corresponds to a slope of $1$ on log-log scale. 
	The transition is discontinuous in $G$ at $\gamma = \gamma_C$
	{\bf (b)} The Non-affinity parameter near the  onset of the transition for $p_0=3.84$ and different \ka. Non-affine cell displacements at below{\bf (c)} and at{\bf (d)} the onset of the transition. 
	}
	\label{fig_strain}
\end{figure}
The  stress-strain relation shown in Fig. \ref{fig_stress-strain}(a)   for a range of values of $p_0$ and  $\kappa_A=0$ reveals three regimes.  For infinitesimal strain the solid responds linearly with modulus $G_0$. In the fluid, $G_0=0$. At intermediate strain ($0< \gamma < 1$) we observe strong stiffening. In particular, the liquid acquires a finite rigidity for $\gamma$ above a critical value $\gamma_C(p_0)$.
At  larger  strains ($\gamma \gtrsim 2$), the  tissue undergoes plastic rearrangements via T1 transitions, resulting in intermittent stick-slip behavior. We define the dynamic yield stress $\sigma_{\text{yield}}(p_0)$ by averaging $\sigma$ in the plastic regime ($2 < \gamma <6$).
 The yield stress is large in a solid tissue  and  decreases as $p_0$ increases, {vanishing at} $p_0 \sim 4.03$ (see  Fig.S1). 
{The main} focus of this work is the stress response in the intermediate region of strain stiffening and strain-induced rigidity,
which is also the regime  most relevant to  experiments\cite{Harris_PNAS_stretch}. We show below that in this regime 
the linear response ($\gamma\rightarrow 0$) cannot predict what happens at finite strain values. 

\paragraph{Shear-induced rigidity transition.} When the unstrained tissue is fluid ($p_0 >p_0^*$), an applied shear strain $\gamma\geq\gamma_C$ yields a finite stress  ~(Fig. \ref{fig_stress-strain}(a)).
The  line  $\gamma_C(p_0)$ where the instantaneous shear modulus  
$
G  \equiv {\partial \sigma}/{\partial \gamma}
$
  vanishes identifies a strain-induced rigidity transition  (Fig.\ref{fig_stress-strain}(b)). In the solid ($p_0< p_0^*$), %
  we observe stiffening for any finite $\gamma$, and $\gamma_C(p_0) = 0 $.  For $p_0 \in [p_0^*,4.03]$, a nonzero value of strain is always required for rigidity and $\gamma_C(p_0)$ grows monotonically with  $p_0$. Beyond $p_0\gtrsim 4 $ 
  { the} tissue {remains} fluid-like regardless of the applied shear strain. 
  This  is consistent with the vanishing of $\sigma_{\text{yield}}$ for $p_0 > 4.03$. 
  The shear stiffening of the liquid  was also reported in recent work on a regular (crystalline) vertex model\cite{ Rastko_2021}, in spring-networks\cite{Merkel_minimal_length_PNAS_2019}{\color{red} and in deformable particle models\cite{corey_stiffening}}.
The mean-field {analysis} below  provides a universal explanation for this behavior.

The nature of the strain-induced  rigidity transition depends on the value of the area stiffness \ka. This is evident in Fig.\ref{fig_strain}(a), where we plot  $G$ near the rigidity  onset  as a function of $\gamma-\gamma_C$.  
At \ka$=0$, the onset of rigidity
is discontinuous.  The jump discontinuity at   $\gamma_C$ remains finite well above $\kappa_A =0$ and becomes vanishingly small and  indistinguishable from a continuous increase in $G$ at $\kappa_A \gtrsim 10^{-3}$. 
For $\gamma<\gamma_C$ the tissue is a marginally rigid solid\cite{moshe2018geometric,Merkel_minimal_length_PNAS_2019} with $G \approx \kappa_A$(Fig.\ref{fig_strain}(a):inset). This is highlighted by the behavior of 
 the fluctuations near the strain-driven rigidity transition, which are quantified  with the non-affinity parameter
$\delta \Gamma = \frac{1}{N \bar{A} \Delta \gamma^2} \langle \left( \delta \mathbf{r}_i -  \delta \mathbf{r}_i^{\text{affine}} \right)^2 \rangle$\cite{Langer_Liu_nonaffinity_1997, DiDonna_Lubensky_nonaffine, Fiber_network_review_Mackintosh}.
Here  $\delta\mathbf{r}_i$ is the displacement of cell $i$ after a strain step and  $ \delta \mathbf{r}_i^{\text{affine}} = \Delta \gamma \ y_i  \ \mathbf{\hat{x}}$ is the affine deformation of the cell located at $\mathbf{r}_i = (x_i, y_i)$.  As shown in Fig.\ref{fig_strain}(b), at low area elasticity ($ \kappa_A \lesssim 10^{-3}$), $\delta \Gamma$ grows monotonically with strain and exhibits a sharp peak at $\gamma_C$, which coincides with  the rigidity transition. At higher $\kappa_A$, there is no pronounced peak in $\delta \Gamma$, indicating a smooth cross-over  from the marginal solid to a rigid solid, rather than a discontinuous  transition.

\begin{figure}[ht]
	\centering
	\includegraphics[width=1\columnwidth]{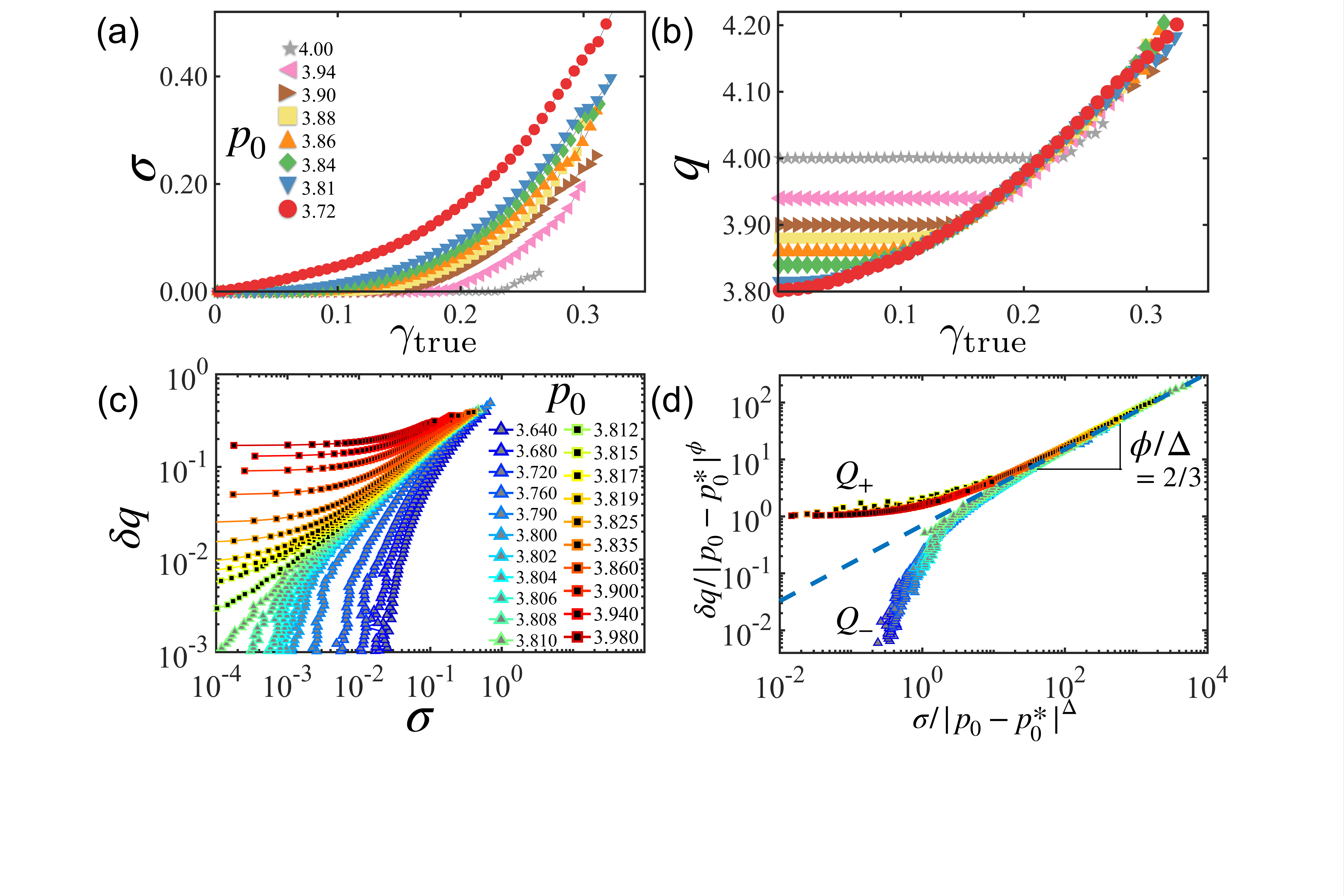}
	\caption{
	{\bf Cell shapes under shear}
	{\bf(a)} A plot of $\sigma$ as a function of \ts for different $p_0$'s spanning the solid and liquid regimes.
	{\bf(b)} The cell shape index $q$ vs the true strain \ts for the same range of $p_0$ as in (a).   
	{\bf(c)} A plot of $\delta q \equiv q - p_0^*$ vs. $\sigma$ for various values of $p_0$ as indicated.  
	{\bf(d)} Replotting of the data  in(c) using the universal scaling ansatz (Eq.\eqref{shape_scaling}). Here $\Delta = 3/2$, $\phi = 1$. All figures are for $\kappa_A=0$. 
	}
	\label{fig_shape}
\end{figure}

\paragraph{Relating mechanical response to cell shape.}
The strain stiffening behavior above $\gamma_C(p_0)$ can be understood in terms of shear-induced changes in the  structural properties of the cellular network. Past work on  vertex models has shown that the observed cell shape index, $q\equiv\langle p/\sqrt{a} \rangle$, is an important metric of the rheological state of the tissue\cite{Park_NMAT_2015,mitchel_ncomm_2020}. We have examined the evolution of this order parameter with applied shear. 
We note, however, that the applied strain $\gamma$ 
does not uniquely define the state of the  tissue due to  plastic events and non-affine deformations. Instead we use the  true strain \ts\cite{gurtin_cont_mech_book} 
to quantify the degree of deformation of the tissue.   \ts  is calculated from the instantaneous deformation tensor of the whole tissue  and therefore  captures the degree  of {\it cumulative}   strain deformation~\cite{SI_text}. 
The motivation for introducing \ts is similar to that behind the fabric tensor in granular materials\cite{Goddard1998} or the recoverable strain in rheology\cite{Recoverable_strain_JOR}. In Fig.\ref{fig_shape}(a,b) we show the stress $\sigma$ and the structural order parameter $q$ as functions of \ts.
It is evident from Fig.\ref{fig_shape}(b) that under shear cell shapes in the fluid stay constant at the \emph{energetically preferred value $p_0$} until the fluid strain-stiffens, while in the solid  $q$ always starts out at the universal value $p_0^*$ and grows quadratically with \ts . 
This behavior is well described by
\be
q =  
\begin{cases}
	p_0, & \gamma_{\text{true}} \le \gamma_C(p_0) \\
	p_0^* +  c \ \gamma_{\text{true}}^2, &  \gamma_{\text{true}} > \gamma_C(p_0). 
\end{cases}
\label{shape_index}
\ee
In the next section, we offer a theoretical derivation of this form.
A similar functional dependence of the observed cell shape on the cell elongation induced by internally generated active stresses  was reported in a recent study of the developing fruit fly\cite{Wang13541}.   

Eq.\eqref{shape_index}  suggests that the quantity $\delta q \equiv q - p_0^*$ can be used as a  \emph{morphological order parameter},  quantifying the deviation of the measured cell shape from the \emph{critical cell shape}.  Moreover, Figs.\ref{fig_shape}(a,b) suggest that the three state variables ($\sigma,\gamma_{\text{true}},\delta q$) are not  independent, and that  any two are sufficient to describe the state of the tissue. Therefore, we eliminate \ts and  plot $\delta q$ as a function $\sigma$ (Fig.\ref{fig_shape}(c)) for a large range of $p_0 \in [3.72, 4]$. 
This plot shows typical hallmarks of a critical point, with qualitatively different behavior above and below $p_0^*$, suggesting  a scaling ansatz 
\be
\delta q  = {\vert p_0-p_0^* \vert}^\phi 
Q_{\pm}\left(\frac{\sigma}{{\vert p_0-p_0^* \vert}^\Delta} \right)\;.
\label{shape_scaling}
\ee
Here $Q_{\pm}(x)$ are the branches of the universal  scaling function for $p_0 > p_0^*$ and $p_0  \le p_0^*$, respectively, with  $x = \sigma/ {\vert p_0-p_0^* \vert}^\Delta$.    This ansatz provides  a nearly perfect collapse of the data (Fig.\ref{fig_shape}(d)), with  $\Delta =3/2$ and $\phi = 1$.  
For   $p_0 > p_0^*$ the behavior is controlled by $Q_{+}(x)$, with $Q_{+}(x)\to{\rm constant}$ for $x  \to 0$, i.e., $\sigma \to 0 $,  implying $\delta q \propto {\vert p_0-p_0^* \vert}^\phi$.
When $p_0 < p_0^*$, the scaling is controlled by $Q_{-}(x)$. In the limit of $\delta q \to 0 $ (i.e., $y = \delta q/ {\vert p_0-p_0^* \vert}^\phi \to 0$), the inverse of $Q_{-}$ tends to a constant, hence  $\sigma \propto {\vert p_0-p_0^* \vert}^\Delta$.  
For $\vert{ p_0-  p_0^* \vert} \to 0$ and $\sigma \gg 0$, the two universal branches merge and  $Q_{+}(x) = Q_{-}(x) = x^{\phi/\Delta}$.

\begin{figure}[htbp]
	\centering
	\includegraphics[width=1\columnwidth]{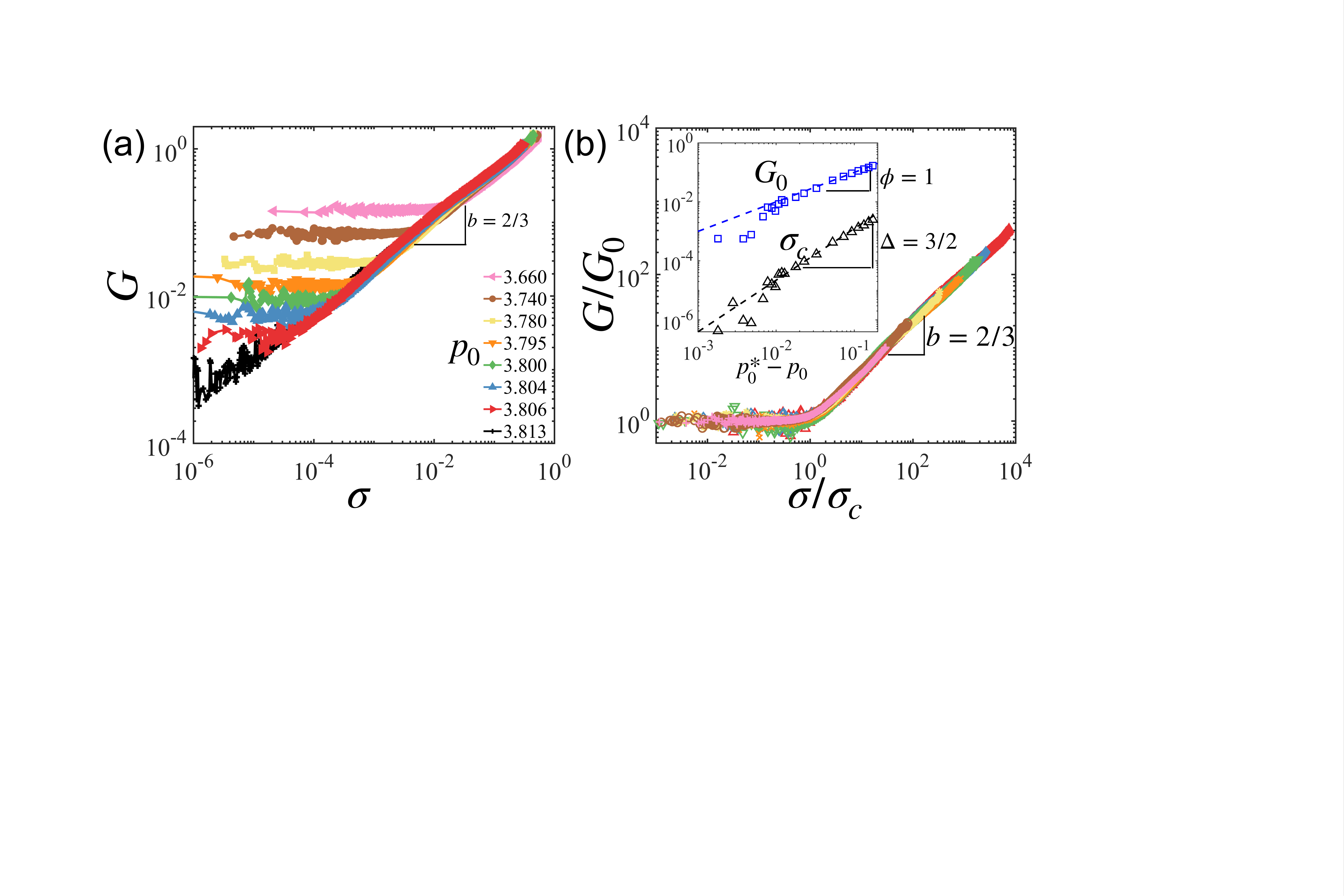}
	\caption{
		(a) The shear modulus $G$ vs. stress $\sigma$ at various  $p_0$ and $\kappa_A=0$. 
		(b) Rescaled $G/G_0$ vs  $\sigma/\sigma_c$ for same set of $p_0$ as in(a).
	}
	\label{fig_nonlinear}
\end{figure}

\paragraph{A nonlinear constitutive equation for sheared tissue.}
In tissues strained beyond  $\gamma_C$  both the stress $\sigma$ (Fig.\ref{fig_stress-strain}a) and the shear modulus $G$ (Fig.\ref{fig_strain}a) are nonlinear functions of the applied strain $\gamma$.
 To quantify the nonlinearity and extract a constitutive equation for the tissue,
we use  $\sigma$, instead of $\gamma$, as a state variable  and plot $G$ as a function of $\sigma$ in Fig.\ref{fig_nonlinear}a for various  $p_0\in[3.66, 3.81]$.  
At small $\sigma$,  $G=G_0$ is independent of $\sigma$, corresponding to  linear elasticity. At higher stress, the elastic response is nonlinear  and $G\propto \left( \sigma/\sigma_c \right)^b$, with  
 $b = 2/3 $. Using $ G = \partial{\sigma}/\partial\gamma$ and eliminating $G$, this yields  a constitutive relation  $\sigma \propto \gamma^{\frac{1}{1-b}} = \gamma^{3}$.
The linear and nonlinear regimes are separated by a critical stress threshold $\sigma_c(p_0)\sim
\vert p_0-p_0^* \vert$.  The linear-response modulus $G_0$ also shows power-law scaling in $\vert p_0-p_0^* \vert$\cite{bi_nphys_2015,yan_bi_prx}. This behavior can be summarized through   a  scaling ansatz to describe the behavior of $G$ in the vicinity of the critical point $p_0^*$
\be
G = {\vert p_0-p_0^* \vert}^\phi \ \mathcal{G} \left(\frac{\sigma}{{\vert p_0-p_0^* \vert}^\Delta} \right).
\label{G_scaling}
\ee
This form provides an excellent collapse of all our data onto a single master curve independent of $p_0$ (Fig.\ref{fig_nonlinear}b).  
From the scaling collapse we obtain  $G_0 \propto {\vert p_0-p_0^* \vert}^\phi $ and $\sigma_c \propto {\vert p_0-p_0^* \vert}^\Delta$, where $\Delta = 3/2 $ and $\phi = 1 $. Crucially, the stress-stiffening scaling collapse (Eq.\eqref{G_scaling}) is directly related to the cell shape-stress scaling relation (Eq.\eqref{shape_scaling}) as $b = \phi/\Delta$.  

\paragraph{Mean-field model of a sheared tissue.}
To gain a theoretical understanding of the strain-driven rigidity and emergence of nonlinear elasticity, we examine a mean-field theory (MFT) formulation  of the vertex model\cite{Czajkowski_hydro_2018,Hernandez_Marchetti,hernandez2021geometric}.  Neglecting cell-cell correlations, we consider the shear deformation of a \emph{single} $n$-sided polygonal cell.
Under affine deformations, the vertex coordinates of a polygon transform according to $\mathbf{R}' = \hat{D} \mathbf{R}$, where $\hat{D}$ is the deformation  tensor given by 
$\hat{D} 
=
\begin{psmallmatrix}
	D_{xx} & D_{xy} \\
	D_{yx} & D_{yy}
\end{psmallmatrix}$. 
We neglect in  Eq.\eqref{vm_energy} the contribution from cell area which is typically small compared to the perimeter term and {examine area-preserving affine deformations with}  $\det \hat{D} = 1$. For  simple shear  $D_{yx}=0$ and 
$D_{yy} =1/D_{xx}$, leaving  only   $D_{xx}$ and $D_{xy}$ as independent components of $\hat{D}$.

\begin{figure}[htbp]
	\centering
	\includegraphics[width=1\columnwidth]{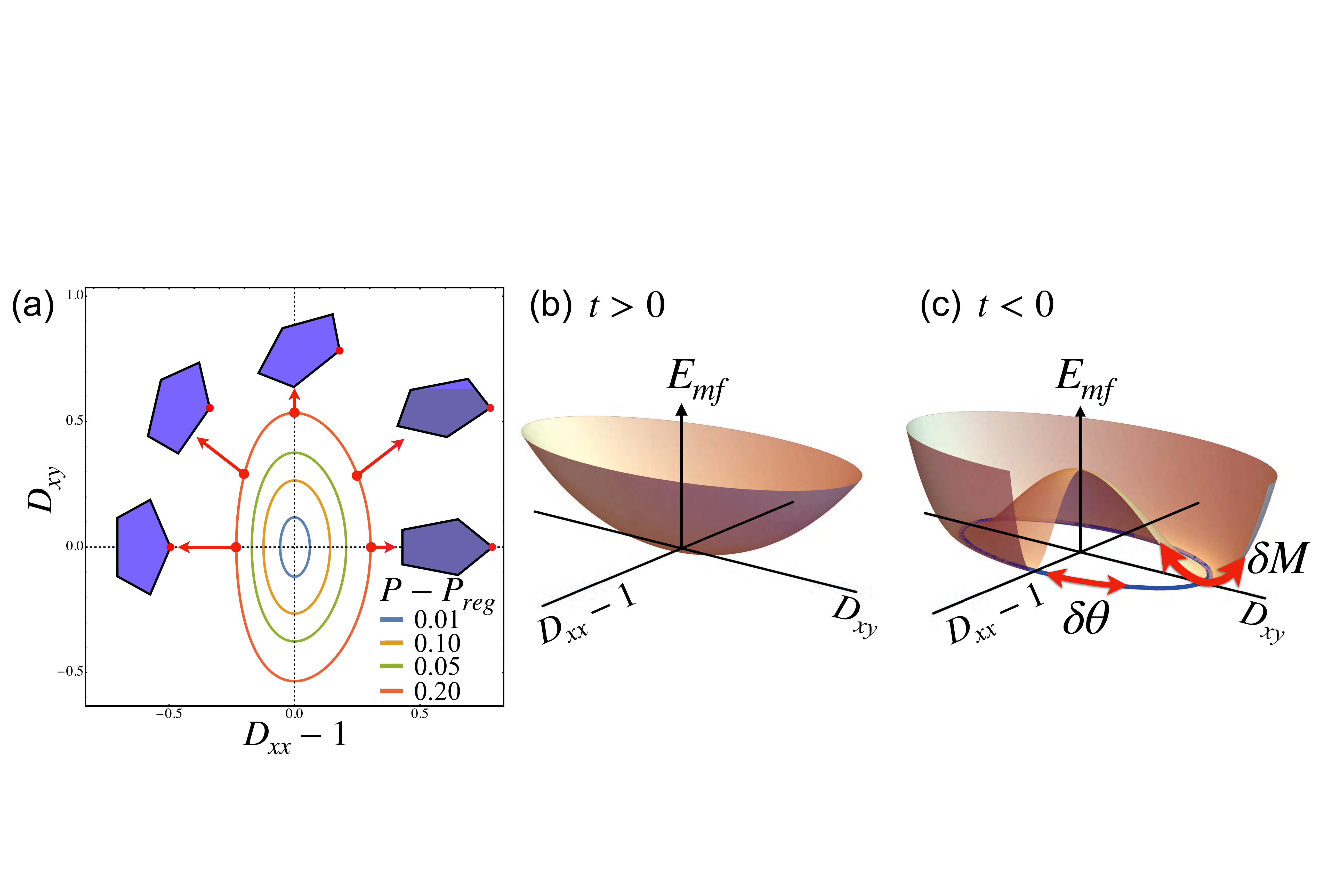}
	\caption{
	(a) When the perimeter of a polygon is larger than that of its regular  counterpart, deformations can lead to a family of isoperimetric shapes defined by the contours shown for a  5-sided polygon. 
	(b) The mean-field energy as a function of ($D_{xx}-1,D_{xy}$) for $t>0$  has a single ground state.
	(c) The mean-field energy  as a function of ($D_{xx}-1,D_{xy}$) for $t<0$  has degenerate ground states which are connected by Goldstone modes along $\delta \theta$ .
	}
	\label{fig_mft}
\end{figure}

The perimeter of a deformed polygon can then be expressed in terms of the components of $\hat{D}$. 
For example the  perimeter of   a quadrilateral ($n=4$) is  given by 
\be
P = \sqrt{2} 
\left[ 
\sqrt{D_{xx}^{-2} + (D_{xx}-D_{xy})^2} + \sqrt{D_{xx}^{-2}  + (D_{xx}+D_{xy})^2}
\right].
\label{p_quad}
\ee
 Expressions for any deformed n-gon are given in the SI\cite{SI_text}. For any $n$, the isoperimetric inequality defines the perimeters compatible with a fixed area as 
 $P>P_{reg}$, where $P_{reg}$ is the perimeter of a regular polygon with unit area (e.g., $P_{reg} = 4$ for $n=4$). The condition $P(D_{xx},D_{xy})\geq P_{reg}$, with $P(D_{xx},D_{xy})$ given by Eq.~\eqref{p_quad}, then defines a manifold in the  $(D_{xx}, D_{xy})$ plane where there exist deformed polygons that statisfy the isoperimetric constraint (Fig.\ref{fig_mft}(a)). 
The maximum value of $D_{xy}$ along the isoperimetric contour defines the   largest simple shear $D_{xy}^{\text{max}}$ that a cell can sustain {by changing its shape, while maintaining its area and   perimeter constant. This value is $\gamma=
\gamma_C = D_{xy}^{\text{max}}  \propto (p_0 - p_0^*)^{1/2}$ and}  precisely corresponds to the location of the strain-driven rigidity  $\gamma=\gamma_C$ in the simulations. 
 The exponent $1/2$ is in excellent agreement  with the $\gamma_C$ scaling in the vicinity of $p_0^*$, shown in Fig.\ref{fig_stress-strain}:inset.

The isoperimetric contours are centered at ($D_{xx}=1, D_{xy}=0$) and well approximated by an ellipse for small $P-P_{reg}$. We introduce polar coordinates  with radius $M(\theta)$ and polar angle $\theta$: $D_{xx}-1 = M(\theta) \cos \theta$ and  $D_{xy}  =M(\theta)  \sin \theta$ and expand Eq.\eqref{p_quad} to $\mathcal{O}(M^2)$ to give  ( see SI\cite{SI_text}) 
\be
P \approx P_{reg} + \frac{15}{32} P_{reg} \left[1+ \frac{3}{5} \cos(2\theta) \right]  \ M(\theta)^2. 
\label{p_approx}
\ee  

Using Eq.\eqref{p_approx}, we  rewrite the   vertex model energy (Eq.\eqref{vm_energy})  to obtain a Landau-type energy
\be
E_{mf}= \frac{1}{2} \ t \ \alpha  \  m(\theta,M)^2 + \frac{1}{4} \beta m(\theta,M)^4,
\label{E_mf}
\ee
where  $m(\theta,M) = \left[1+\frac{3}{5} \cos(2\theta)\right]^{1/2} M$ is the order parameter, $\alpha = (60/32) {p_0^*}^2, \ \beta = (30/32) {p_0^*}^2$ are positive constants, and $t = (p_0^* - p_0)/p_0^*$ controls the distance to a continuous phase transition in $m(\theta,M)$. For  $t>0$, $E_{mf}$ has a single minimum at $m^*=0$ (Fig.\ref{fig_mft}b), corresponding to the rigid state. When $t<0$, the minimum $m^*(\theta,M)$ corresponds to the {isoperimetrically degenerate liquid state}. In the energy landscape these states are connected by a Goldstone mode (Fig.\ref{fig_mft}c).

The MFT explains the origin of the nonlinear elasticity. For  $t>0$, $E_{\text{mf}}$ has a single minimum at $m^*=0$ (corresponding to an undeformed solid state) and deformations away from it can be calculated  
using Eq.\eqref{E_mf} 
\be
\begin{split}
\sigma  &= \partial E_{mf}/\partial m = \alpha  \ t m + \beta m^3 \\
G &= \partial^2 E_{mf}/\partial m^2 = \alpha  \ t  + 3\beta m^2.
\end{split} 
\label{mf_mech}
\ee
For small $m$ we recover linear elasticity with $G_0 = \alpha \ t \propto (p_0^* - p_0)$.  At large $m$ the response is nonlinear, with $G \propto \sigma^{2/3}$. The cross-over stress between the two regimes can be calculated:  $\sigma_c = 2\beta \alpha^{3/2} t^{3/2}  \propto (p_0^* - p_0)^{3/2}$. These predictions are in excellent agreement with simulations results. 

We have used a vertex model to study the nonlinear  response of a tissue to shear. Using simulations and MFT, we showed that a tissue that is liquid when unstrained stiffens upon shear. 
Liquid-solid transitions in VM of biological tissues are driven by geometric frustration and active mechanisms. Recent work by some of us~\cite{hernandez2021geometric} showed that geometric incompatibility controls the response to infinitesimal deformations, providing the underlying unifying mechanism for rigidity in a broad class of underconstrained systems.  The present work additionally incorporates active processes that mediate plastic response. Plasticity dominates at higher strains and is likely to underlie the rheology of real tissue.  Both works use a MFT to highlight the geometric origin of the degeneracy of the liquid ground state.  The same MFT is extended here to investigate the response to deformations.  While a Voronoi-based model is used, we have observed the same quantitative behavior using a vertex-based model and the results are independent of the model implementation.

Finally, it was shown in Ref. ~\cite{hernandez2021geometric} that at the critical point the VM shares many of the properties of odd elasticity~\cite{Scheibner_nature_2020} - for instance, spontaneous shear upon uniaxial extension - although this behavior arises from geometry, not from an energy input at the microscale. Exploring the response to deformations other than simple shear and the possible connections with odd elasticity is an important direction for future work.


\begin{acknowledgments}  
The authors thank Mark Bowick, Michael Moshe and Arthur Hernandez for illuminating discussions.
This work was supported in part by  the Northeastern University TIER 1 Grant (J.H. and D.B.),
NSF  DMR-2046683 (J.H. and D.B.), DMR-2041459 (M.C.M.),  PHY-1748958 (D.B. and M.C.M.),  the Center for Theoretical Biological Physics NSF PHY-2019745 (J.H. and D.B.) and Mathworks Microgrants. 
We  acknowledge the support of the Northeastern University Discovery Cluster. 
This project has received funding from the European Research Council (ERC) under the European Union’s Horizon 2020 research and innovation programme (grant agreement No. 885146); and the SOFI CDT, Durham University, EPSRC (EP/L015536/1).
\end{acknowledgments}

\bibliography{shear_paper.bib}

\end{document}